\documentclass[
   10pt,
    a4paper,
    iopart,
    twocolumn,
    bibnotes,
    showpacs,
    superscriptaddress,
    groupedaddress
]{revtex4}

\usepackage[mathscr]{euscript}
\usepackage{bm}
\usepackage{array}
\usepackage{graphicx}
\usepackage{amssymb}
\usepackage{amsmath}
\usepackage{fullpage}
\usepackage{hyperref}
\usepackage{paralist}
\usepackage{subfigure}
\usepackage{float}
\hypersetup{
    colorlinks   = true,
     citecolor    = blue,
     linkcolor    = blue,
     citecolor    = blue,
     filecolor     = blue,
     urlcolor	      =	blue  
}

\setcitestyle{square}
\newcommand{\bs}{\boldsymbol}
\newcommand{\beq}{\begin{equation}}
\newcommand{\eeq}{\end{equation}}
\newcommand{\bra}[1]{\langle #1 |}
\newcommand{\ket}[1]{| #1 \rangle}
\newcommand{\brahket}[3]{\langle #1 | #2 | #3 \rangle}
\newcommand{\braket}[2]{\langle #1 | #2 \rangle}
\newcommand{\ketbra}[2]{\ket{#1}\bra{#2}}

\newcommand{\beqa}{\begin{eqnarray}}
\newcommand{\eeqa}{\end{eqnarray}}

\newcommand{\al}[1]{\begin{align}#1\end{align}}

\newcommand{\Eqref}[1]{Eq.~(\ref{#1})}
\newcommand{\Figref}[1]{Fig.~\ref{#1}}
\newcommand{\Secref}[1]{Sec.~\ref{#1}}

\begin{document}
\title{STM contrast of a CO dimer on a Cu(1 1 1) surface: a wave-function analysis}
\date{\today}

\author{Alexander \surname{Gustafsson}}
\email{alexander.gustafsson@lnu.se}
\affiliation{Department of Physics and Electrical Engineering, Linnaeus University, 391 82 Kalmar, Sweden}
\author{Magnus \surname{Paulsson}}
\affiliation{Department of Physics and Electrical Engineering, Linnaeus University, 391 82 Kalmar, Sweden}

\begin{abstract}
We present a method used to intuitively interpret the STM contrast by investigating individual wave functions originating from the substrate and tip side. We use localized basis orbital density functional theory, and propagate the wave functions into the vacuum region at a real-space grid, including averaging over the lateral reciprocal space. Optimization by means of the method of Lagrange multipliers is implemented to perform a unitary transformation of the wave functions in the middle of the vacuum region. The method enables (i) reduction of the number of contributing tip-substrate wave function combinations used in the corresponding transmission matrix, and (ii) to bundle up wave functions with similar symmetry in the lateral plane, so that (iii) an intuitive understanding of the STM contrast can be achieved. The theory is applied to a CO dimer adsorbed on a Cu(1 1 1) surface scanned by a single-atom Cu tip, whose STM image is discussed in detail by the outlined method.
\end{abstract}

\pacs{68.37.Ef, 33.20.Tp, 68.35.Ja, 68.43.Pq}
\maketitle

\section{Introduction}
Scanning tunneling microscopy (STM) is a mature technique to reveal atomic structures on surfaces. In addition to images of atomic arrangements, detailed measurements can be made of molecular orbitals \cite{Repp2005,Gross2011}, electronic structure, vibrational \cite{Stipe1998} and magnetic excitations \cite{Hirjibehedin2007}. With atomic force microscopy (AFM), the interaction between tip and sample is characterized, which allows for examination of the tip-apex structure \cite{Ternes2008,Emmrich2015prl,Welker2012,Emmrich2015}. The tip-apex structure may strongly influence the tunneling process, both when making a quantitative comparison between experimental and theoretical STM images \cite{Gustafsson2017}, and when investigating the inelastic tunneling signal from an adsorbate species \cite{Okabayashi2016}. 

To model STM images, the approximation methods of Bardeen \cite{Bardeen1961} and Tersoff-Hamann \cite{Tersoff1985} have been widely used. The Tersoff-Hamann approach, which can be derived from the Bardeen's approximation with an $s$-wave tip \cite{Hofer2003}, has provided an intuitive understanding of STM experiments with non-functionalized STM tips \cite{Persson2002}. For CO-functionalized STM, the Bardeen method \cite{Paz2006,Rossen2013,Bocquet1996,Teobaldi2007,Zhang2014}, the Chen's derivative rule \cite{Chen1990, Mandi2015, Mandi2015b}, and Landauer-based Green's function methods \cite{Cerda1997a,Cerda1997b} include the effects of more complicated tip states. 

Our previous work on this topic \cite{Gustafsson2016,Gustafsson2017} concerns first-principles modeling based on Bardeen's approximation, using localized-basis DFT. The wave functions close to the atoms were calculated in a localized basis set using Green's function techniques, whereafter they were propagated into the vacuum region in real space by utilizing the total DFT potential. We concluded that calculations in the $\Gamma$ point ($\bold{k}=\bold{0}$) may, for many systems, qualitatively reproduce the STM contrast of the \textbf{k} averaged calculations, whereas a quantitative comparison to experiments always seems to require averaging over the lateral reciprocal space \cite{Gustafsson2017}. 

In this paper we further develop the theoretical method focusing on providing a simpler interpretations of the STM contrast. The numerous wave functions given by DFT in the vacuum region obscures the interpretation, and here we have developed a unitary transformation to find the dominating tip- and substrate wave function combinations (henceforth denoted tip-sub combinations) that give the largest contribution to a specific STM image. We also present a simple formula, relating the real-space tip- and substrate wave functions to the transmission probability. To illustrate the method, we analyze the STM contrast of a CO dimer \cite{Heinrich2002,Zophel1999,Niemi2004} adsorbed on a Cu(111) surface scanned by a single-atom Cu tip. 

\section{Methodology}
\subsection{Brief summary of the previous work}
\begin{figure}[]
	\includegraphics[width=\columnwidth]{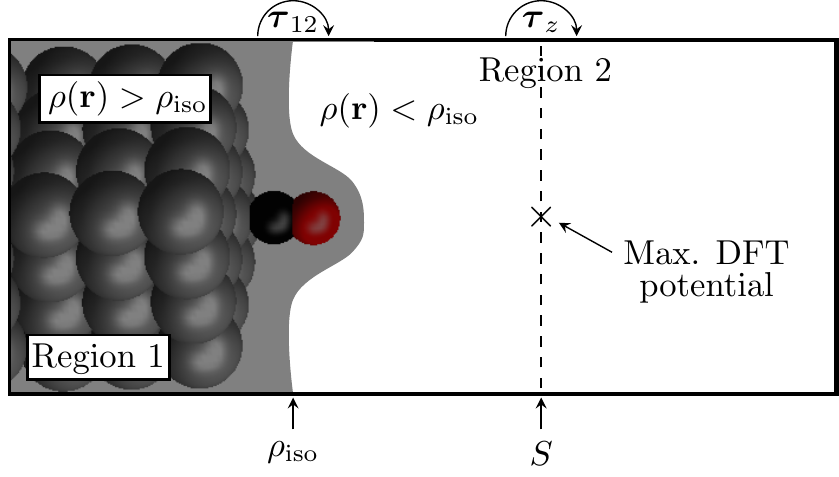}
	\caption{Illustrative description of some of the quantities used for calculation of the substrate wave functions, $\varphi_s^k(\mathbf{r})$, in region 2. The localized-basis wave functions, $\psi_{s}^k(\mathbf{r})$, in region 1 serve as boundary conditions at the electron charge density isosurface, $\rho_{\textrm{iso}}$, and are propagated into region 2 in real space. To the right of the separation plane, $S$, the potential is the average of the potential at $S$, in order to simulate a constant vacuum potential far from the surface. The totally reflected tip wave functions, $\widetilde{\varphi}_t^k(\mathbf{r})$, are calculated similarly, where the potential at the separation plane is modeled by an impenetrable barrier.\label{fig1}}
\end{figure}

In our previous work \cite{Gustafsson2016,Gustafsson2017}, we start by a conventional calculation of the localized-basis wave functions that originate from Bloch states in respective lead \cite{Paulsson2007}. These wave functions are only accurate close to the atoms due to the finite range of the basis orbitals. We therefore propagate these wave functions into the vacuum gap. From the DFT electron charge density, $\rho(\mathbf{r})$, we define a density isosurface, $\rho_{\mathrm{iso}}$, at which these wave functions serve as boundary conditions. A finite-difference (FD) Hamiltonian is thereafter constructed for the device region, which contains the total DFT potential, and a discrete Laplacian.
The device region contains the surface atomic layers of the substrate and the tip, as well as intermediate adsorbates, tip structures, and the vacuum gap. The elements of the FD Hamiltonian are reordered according the real-space positions of the charge density isosurface. This enables a Hamiltonian inside (outside) the isosurface, $\mathbf{H}_1$ ($\mathbf{H}_2$), as well as the coupling matrix, $\bs{\tau}_{12}^{(\dagger)}$, between these regions, c.f., \Figref{fig1}. The lattice plane for which the maximum DFT potential occurs is used as a separation plane, $S$, between the substrate and tip slabs. In order to simulate an isolated substrate slab, the potential further away from this plane is set as the average potential at $S$. The wave functions at the separation plane, at a specific $\mathbf{k}$ point, $k$, are calculated by solving the sparse linear system of equations,
\beq
(\varepsilon_\mathrm{F}\mathbf{I}-\mathbf{H}_2^k)\cdot\varphi_n^k(\mathbf{r})=\bs{\tau}_{12}^\dagger\cdot\psi_n^k(\mathbf{r}),
\eeq
where $\varepsilon_\mathrm{F}$ is the Fermi energy, $\mathbf{H}_2^k$ is the vacuum Hamiltonian (region 2 in \Figref{fig1}), $\varphi_n^k(\mathbf{r})$ is the $n$th vacuum wave function, $\bs{\tau}_{12}^\dagger$ is the coupling between the two regions, and $\psi_n^k(\mathbf{r})$ is the $n$th localized-basis wave function in region 1, i.e., where it is accurate. When using the Bardeen's approximation \cite{Bardeen1961} for the conductance, the wave functions from both sides are calculated similarly, whereafter the wave functions and their derivatives are used for calculation of the conductance. However, as will be further discussed below, the calculation of the tip wave functions is modified, as we adopt a slightly different approach for obtaining the conductance in this paper.

Similarly to our previous work, bias-voltage dependence is not considered in the theory presented below. Extending the present model to include it is straightforward. Firstly, the voltage drop over the vacuum gap needs to be modeled either by a self-consistent DFT calculation \cite{Brandbyge2002}, or introducing the voltage drop in the vacuum gap by hand. Secondly, the $I$-$V$ characteristic can either be explicitly calculated as an integral over energy, or d$I$/d$V$ maps calculated from the wave functions at the chemical potentials of the leads. The unitary transformation discussed below will still be relevant for the d$I$/d$V$ map, but not for the full energy dependent case. Furthermore, for bias-voltage dependence, the DFT bandgap underestimation has to be considered and possibly corrected for \cite{Perdew1985}. 

\subsection{Transmission coefficient}
A handy feature of the simple finite difference approximation, is that the coupling strength operator (matrix), $\tau_z$, between individual lattice planes along the direction of transport ($\hat{z}$) is a simple diagonal matrix \footnote{In contrast to localized-basis calculations, where the coupling strength depends on the tip position, which therefore varies when the tip is shifted laterally over the substrate.}. That is, $\tau_z=-a_z^{-2}\bf{I}$ (in Rydberg atomic units), where $a_z$ is the slice spacing in real space (see \Figref{fig1}). This means that one may exploit the ordinary Green's function formalism to provide an accurate description of transmission probability, that gives the same results as the Bardeen formula. The transmission coefficient for all tip-sub combinations in a specific \textbf{k} point, $k$, may then be expressed as
\beq
T^k_{\textrm{tot}}=4\pi^2\sum_{st}\big|\bra{\varphi_s^k}\tau_z^\dagger \ket{\widetilde{\varphi}_t^k}\big|^2\label{eqtransGF}.
\eeq
This equation can be derived from the more familiar expression for the transmission coefficient (henceforth omitting the superscript $k$), $T_{\textrm{tot}}=\textrm{Tr}\big[G_d\Gamma_sG_d^\dagger \Gamma_t\big]$, which relates the Green's function of the device region (that is, the substrate wave functions when connected to the tip part), to the isolated tip part by the broadening $\Gamma_t$,
\al{
&\textrm{Tr}\big[G_d\Gamma_sG_d^\dagger \Gamma_t\big]=\textrm{Tr}\big[A_s\Gamma_t\big]\\
&=2\pi\sum_{s}\brahket{\varphi_s}{\Gamma_t}{\varphi_s}=2\pi\sum_{s}\brahket{\varphi_s}{\tau_z^\dagger a_t\tau_z}{\varphi_s}\\
&=(2\pi)^2\sum_{s,t}\brahket{\varphi_s}{\tau_z^\dagger}{\widetilde{\varphi}_t}\brahket{\widetilde{\varphi}_t}{\tau_z}{\varphi_s},
}
from which \Eqref{eqtransGF} follows. In the derivation, the broadening of the electronic states from the tip side, $\Gamma_t$, is expressed by utilizing the partial spectral function, $a_t$, so that $\Gamma_t=\tau_z^\dagger a_t\tau_z$, which in turn is composed by the totally reflected tip wave functions, $a_t=2\pi\sum_t\ketbra{\widetilde{\varphi}_t}{\widetilde{\varphi}_t}$ (see Ref.~\cite{Paulsson2007} for further details). This means that the total transmission is proportional to the sum of the squares of the overlap of all tip-sub combinations, due to the utilized FD approximation for the wave functions.

An important note is that the substrate wave functions used in the Bardeen formula and in  \Eqref{eqtransGF} are exactly the same, whereas the tip wave functions in \Eqref{eqtransGF}, $\widetilde{\varphi}_t$, are totally reflected at the separation plane, in contrast to the Bardeen formula. The isolation of the tip side is modeled by introducing an impenetrable barrier at the separation plane when propagating the wave functions from the tip side. This means that the totally reflected tip wave functions, in all essentials, are identical to the normally propagated ones, apart from the amplitude, which depends on the lattice constant $a_z$. This subtle difference in the calculation of the wave functions enables usage of \Eqref{eqtransGF} instead of the Bardeen formula, and the two methods give identical results with same memory consumption in the same CPU time. However, calculation of the derivatives of the wave functions (needed in Bardeen's approximation) is not necessary in \Eqref{eqtransGF}, and the suggested formula provides shorter and simpler derivations upon considering the forthcoming unitary transformation of the wave functions.

\subsection{Unitary transformation of wave functions}
\label{secUnitaryTransformation}
Below we have approximately 1000 tip-sub combinations, and to interpret the calculated STM contrast we describe a simple method to perform a unitary transformation of the wave functions. The main idea is to maximize the amplitude of the wave functions on the separation plane, i.e., maximize $\brahket{\varphi_s^i}{P}{\varphi_s^j}$ where $P$ is the projection on the separation plane  (and similarly for the tip wave functions), so that the important wave functions from each side of the vacuum region become ordered by their amplitude, which in turn approximately correlates to their importance as tunneling channels. Maximization with respect to the conductance, i.e., $|\brahket{\varphi_s^i}{\tau_z}{\widetilde{\varphi}_t^j}|^2$, may be appropriate when considering a fixed tip position, where each tip-sub combination gives a single-valued conductance. However, since we consider a tip scanning over the whole substrate, the former maximization is more relevant in the present context.

The $i$th unitary transformed wave function can therefore be written as a linear combination of the accessible propagated wave functions,
\beq
\ket{\psi_i}=\sum_{j=1}^nc_{ij}\ket{\varphi_j}\label{eqTransf},
\eeq
subjected to the the constraint for the expansion coefficients, $\sum_{j=1}^{n}|c_{ij}|^2=1$, where $n$ is the number of considered wave functions from each side. By means of the Lagrange multiplier method \cite{Arfken2012,Bertsekas1982} within the principal component analysis \cite{Pearson1901,Jolliffe2002}, the equation to solve (for each $j$) reads,
\beq
\frac{\partial}{\partial {c_s^{i}}^*}\bigg[\sum_{ij}\brahket{\varphi_s^i}{{c_s^{i}}^* P c_s^j}{\varphi_s^j}-\lambda_s^j\big(\sum_i{c_s^{i}}^*c_s^j-1\big)\bigg]=0,
\eeq 
upon considering transformation of the substrate wave functions. This equation reduces to a standard eigenvalue problem,
\beq
\left(\mathbf{S}_s-\lambda_s^j\mathbf{I}\right)\cdot\vec{c}_s^{~j}=0\label{eqevproblem},
\eeq
where $\mathbf{S}_s\equiv[[\brahket{\varphi_s^i}{P}{\varphi_s^j}]]$ is the matrix containing all overlap integrals on the separation surface between individual wave functions. The desired expansion coefficients, $c_s^{ij}$, are therefore directly obtained by diagonalization of $\mathbf{S}_s$, \Eqref{eqevproblem}, so that the numerical implementation is trivial. This procedure is performed similarly for the totally reflected wave functions from the tip side, so that a different transformation matrix is obtained for these wave functions, unless the substrate and tip slabs are identical.

An important feature of the outlined method is that the sum of the overlaps of all tip-sub combinations squared, \Eqref{eqtransGF}, is invariant under the unitary transformation, \Eqref{eqTransf}. This property may conveniently be proved backwards, by expansion of the unitary transformed substrate wave functions,
\begin{widetext}
\al{
&\sum_{ij}\big|\braket{\psi_t^j}{\psi_s^i}\big|^2=\sum_{ij}\braket{\psi_t^j}{\psi_s^i}\braket{\psi_s^i}{\psi_t^j}=\sum_{j}\bra{\psi_t^j}\sum_i\bigg[\sum_{k}c_s^{ik}\ket{\varphi_s^k}\sum_{k'}\bra{\varphi_s^{k'}}{c_s^{ik'}}^*\bigg]\ket{\psi_t^j}\\
&=\sum_j\bra{\psi_t^j}\bigg[\sum_{ikk'}{c_s^{ik'}}^*c_s^{ik}\ket{\varphi_s^k}\bra{\varphi_{s}^{k'}}\bigg]\ket{\psi_t^j}\label{eqMixterms}=\sum_j\bra{\psi_t^j}\bigg[\sum_k\ket{\varphi_s^k}\bra{\varphi_{s}^{k}}\bigg]\ket{\psi_t^j}=\sum_{jk}\big|\braket{\psi_t^j}{\varphi_s^k}\big|^2,
}
\end{widetext}
where the orthogonality of the eigenvectors, $\braket{c_s^{ik}}{c_s^{ik'}}=\delta_{kk'}$, is exploited, so that the mix terms vanish. The same procedure applied to the unitary transformed tip wave functions, $\ket{\psi_t^j}$, completes the proof. This shows that the total transmission, \Eqref{eqtransGF}, (for a specific $\mathbf{k}$ point) remains invariant under a unitary transformation, and ultimately that the \textbf{k} averaged transmission coefficient is unaffected by such a transformation. Hence, the unitary transformation matrices, $\bm{\mathscr{U}}_{s(t)}$, are found, such that $\ket{\vec{\psi}_{s(t)}}=\bm{\mathscr{U}}_{s(t)}\,\ket{\vec{\varphi}_{s(t)}}$, leaving the physical properties unchanged.

This procedure often seems to result in that wave functions of the same lateral symmetry, i.e., highly correlated variables, are collected and thereby reduce the dimensionality of the problem. This means that a smaller subset of wave functions are needed in the transmission matrix to obtain an accurate description of the STM image.

\subsection{Computational details}
We use the \textsc{Siesta} \cite{Soler2002}  DFT code to geometry optimize a slab consisting of eight Cu layers, where each layer has $6\times6$ Cu atoms with a nearest-neighbour distance of 2.57 \AA, so that the lateral cell dimensions are $15.4\times15.4$ \AA. In the calculations presented below we use a $7.5$~\AA~core-core distance along $\hat{z}$ between the very apex of the Cu tip and the oxygen atoms (see \Figref{fig2}), which assures a negligible interaction between the tip and the substrate \cite{Gustafsson2016}, so that the substrate wave functions are unaffected when simulating the scanning of the STM tip over the surface.

The  \textsc{Siesta} calculations are performed using the Perdew-Burke-Ernzerhof (PBE) parametrization of the generalized gradient approximation (GGA) exchange-correlation functional \cite{Perdew1996}, double- (single-) zeta polarized basis set for C, O (Cu) atoms, a 200 Ry real space mesh cutoff, and $4\times4$ $\bold{k}$ points in the surface plane. The CO molecules are adsorbed on two adjacent top sites of the otherwise clean Cu(111) substrate surface, and on the opposite side of the vacuum region a pyramid tip consisting of four Cu atoms are attached to a similar surface, see \Figref{fig2}. The geometry optimization concerns the adsorbate species, the two top substrate layers, and the tip atoms (forces less than 0.04 eV/\AA). Nine additional Cu layers are thereafter added (three at the bottom of the substrate, and six above the tip slab), and the wave functions are calculated by \textsc{Transiesta} \cite{Brandbyge2002}, \textsc{Inelastica} \cite{frederiksen2007}, and our recent STM model \cite{Gustafsson2016,Gustafsson2017}. 

The \textbf{k} grid used in the STM calculation consists of $11\times11$ \textbf{k} points that are homogeneously distributed in reciprocal space, and shifted so that an odd number of \textbf{k} points includes the $\Gamma$ point. In the following analysis, the wave functions obtained from a $\Gamma$-point STM calculation are discussed. It is therefore important that the considered $\textbf{k}$ averaged STM image agrees qualitatively with the $\Gamma$-point calculation, unless there is interest in investigating several $\textbf{k}$ points. 

\section{Results}
\subsection{Calculated STM contrast of a CO dimer}

\begin{figure}[t!]
	\includegraphics[width=\columnwidth]{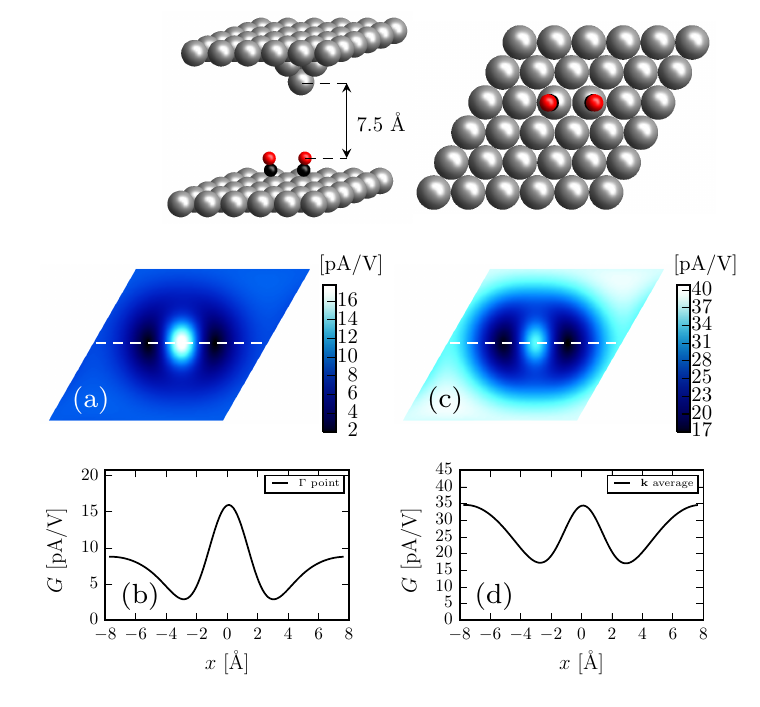}
	\caption{The top panel shows the relaxed geometry of two CO molecules adsorbed on adjacent top sites of a Cu(111) surface, as well as the four-atom pyramid Cu(111) tip, which apex consists of a single Cu atom. (a) shows the $\Gamma$-point STM image, and (b) shows the cross-section conductance along the dashed white line in (a). (c) and (d) are the \textbf{k}-averaged counterparts sampled at an $11\times11$ \textbf{k} grid.\label{fig2}}
\end{figure}

A single CO molecule prefers to stand upright when adsorbed on a top site of a Cu(111) surface, i.e., the C-O bonding axis is perpendicular to the surface. Such an adsorbate exhibits a solid radially symmetric conductance dip when scanned by a pure Cu tip \cite{Gustafsson2016,Gustafsson2017}.

The present CO dimer, i.e., two CO molecules adsorbed on adjacent top sites of a Cu(111) surface, reveals a bright spot centered in a slightly elongated surrounding dip, when scanned in constant-current mode by an $s$-wave tip \cite{Heinrich2002}. A similar feature is also experimentally observed for multiple CO monomers on a Cu(211) surface \cite{Zophel1999}. Furthermore, the CO dimer exhibits slightly tilted bonding angles of the C-O bonding axes, as the oxygen atoms push apart, which has an evident effect of the conductance perpendicular to the surface \cite{Niemi2004}. We confirm this feature [\Figref{fig2}(c)] in constant-height mode, when scanning the molecule by a four-atom pyramidal Cu tip at tip height 7.5~\AA, averaging over a relatively dense ($11\times11$) \textbf{k} grid. The CO bonding length is 1.17~\AA, and the  tilt angles of the C-O bonding axes deviate $(6.6\pm0.5)^\circ$ from standing perpendicular to the Cu(111) surface, in agreement to previous studies \cite{Persson2004}. We have also noticed that the calculated STM contrast for this system comprises a certain dependence on the tip height. By lowering the tip by 1.5~\AA, the central bright spot becomes more pronounced, whereas elevating the tip by 1.5~\AA, lowers the central bright spot. 

\begin{figure}[tbh!]
	\includegraphics[width=\columnwidth]{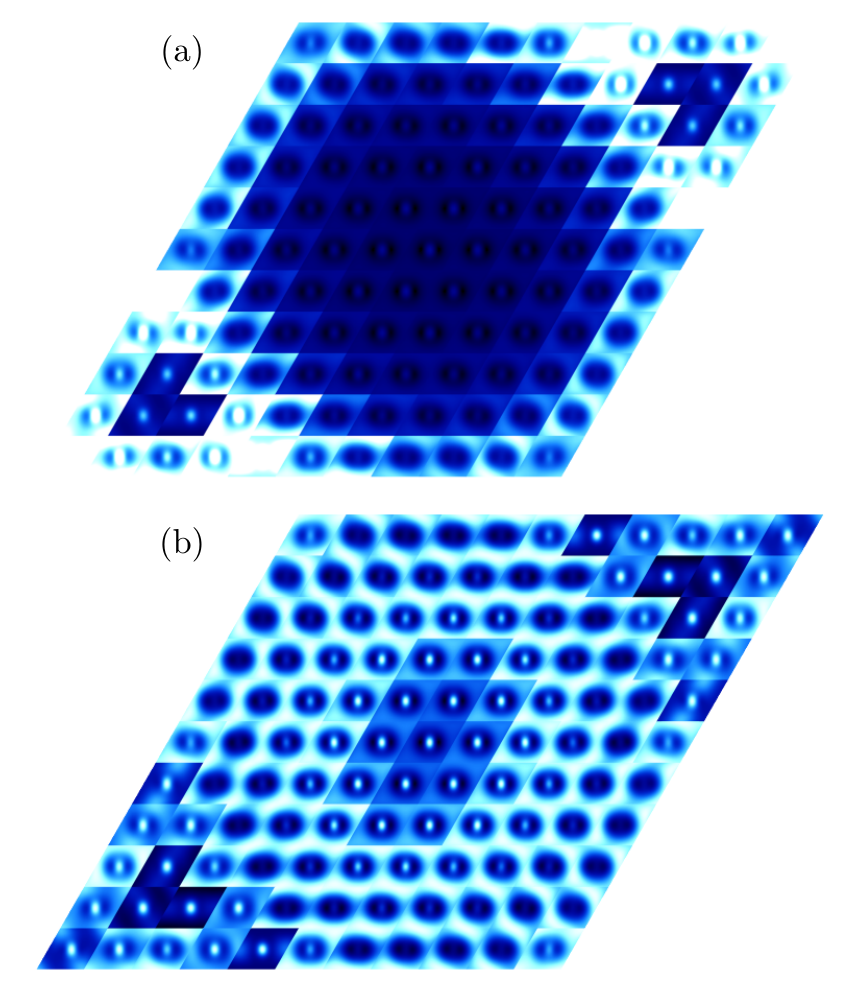}
	\caption{(a) Individual STM images for \textbf{k} points that span the first Brillouin zone in the lateral plane, using the same color scale [bright (dark) means high (low) conductance] for all \textbf{k} points, which highlights their significance for the conductance. (b) Same image as (a), upon imposing individual color scaling for each \textbf{k} point, so that the feature/symmetry for each \textbf{k} point is transparent. The $\Gamma$ point lies in the middle of the figures.\label{fig3}}
\end{figure}

For this, and other systems \cite{Gustafsson2017}, a $\Gamma$-point STM calculation [\Figref{fig2}(a)] qualitatively reproduces the \textbf{k} averaged STM image [\Figref{fig2}(c)], which allows to restrict the forthcoming wave function analysis to the $\Gamma$ point. The main difference between these calculations is an overall increment of conductance, as well as a larger contribution from the Cu substrate surface atoms, when performing averaging in reciprocal space. This feature is visualized in \Figref{fig3}, where the conductance in vicinity of the $\Gamma$ point is smaller than closer to the edges of the surface Brillouin zone. The unit [pA/V] for the conductance, used in \Figref{fig2}, is acquired by conventionally relating the transmission probability to the conductance with the formula $G=G_0T_{\textrm{tot}}\times10^{12}$, where $G_0$ is the conductance quantum, and $T_{\textrm{tot}}$ is the transmission coefficient calculated by \Eqref{eqtransGF}. According to \Figref{fig3}(b), we conclude that also \textbf{k} points further away from the Brillouin zone center give $\Gamma$-point-like STM images. Therefore, by choosing a \textbf{k} grid that excludes the $\Gamma$ point does not have an impact on the \textbf{k} averaged STM contrast. For instance, a coarser \textbf{k} grid ($6\times6$) that excludes the $\Gamma$ point, gives an almost identical STM image compared to the $11\times11$ grid used here.

\subsection{Unitary transformation of wave functions}

\begin{figure*}[tbh!]
	\includegraphics[width=\textwidth]{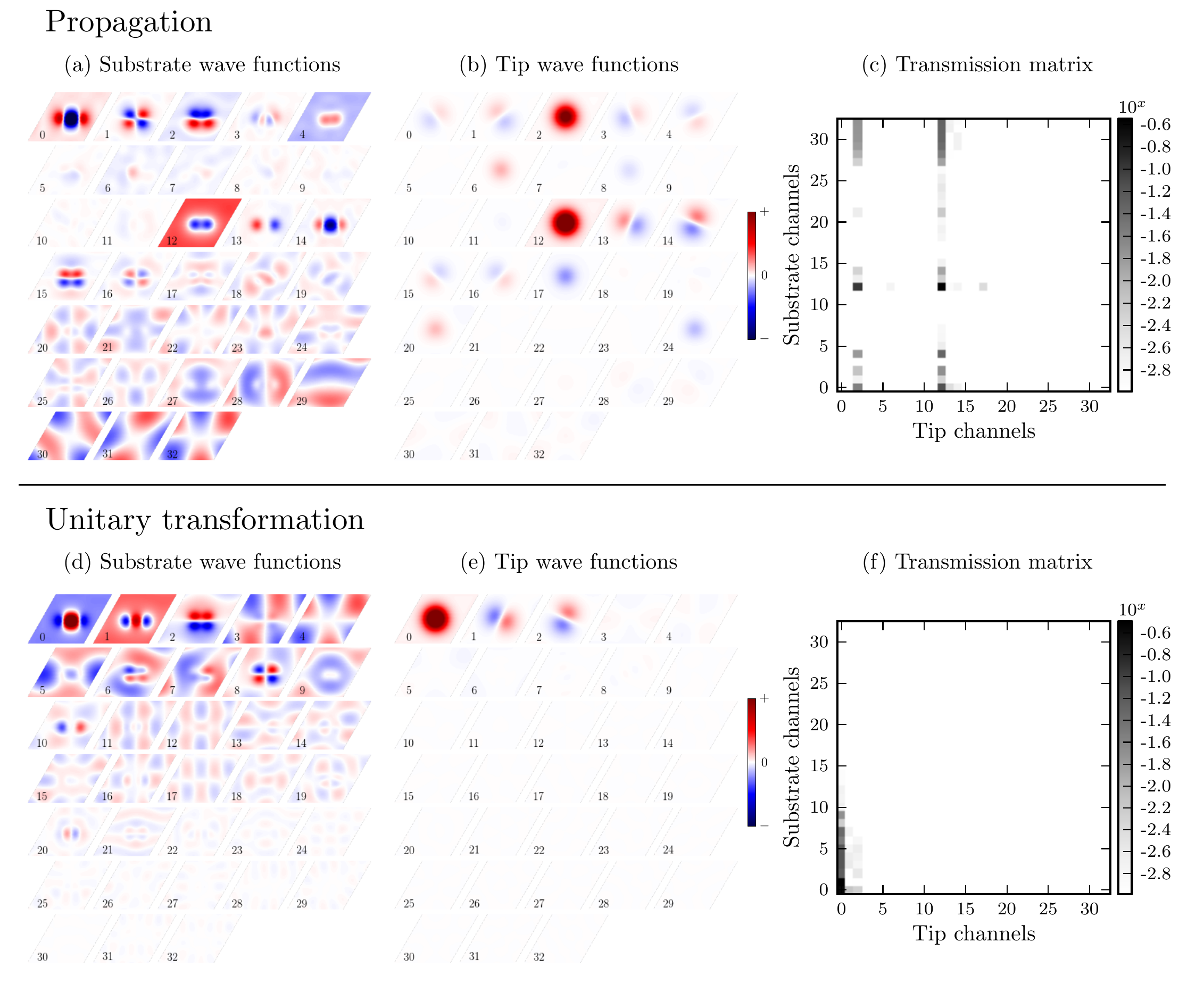}
	\caption{Upper panel shows (a) the normally propagated substrate wave functions, and (b) the tip wave functions (real part only). (c) shows a logarithmically scaled transmission matrix that highlights the important tip-sub combinations, normalized so that the total transmission is unity, and where elements with values below $10^{-3}$ are omitted. The bottom panel shows the corresponding results after performing the unitary transformation of the wave functions in (a) and (b). The plot range is individual for each set of wave functions, and clipped so that the maximum amplitude wave function lies outside the range, in order to also highlight the feature/symmetry of the less significant wave functions, e.g., the $p$ waves from the tip. \label{fig4}}
\end{figure*}

Despite that the considered STM image is already calculated and briefly discussed in reciprocal space, the origin of the STM contrast still needs to be interpreted. Each $\textbf{k}$-point STM image in \Figref{fig2} is a result of approximately 1000 ($33\times33$) individual tip-sub combinations used in \Eqref{eqtransGF}. The significance of these combinations to the STM contrast may be examined directly from the propagated wave functions. However, the unitary transformation of these wave functions, described in \Secref{secUnitaryTransformation}, yields a more transparent understanding of the STM contrast.

In \Figref{fig4}(a) and \Figref{fig4}(b) the real part \footnote{The imaginary part is negligible in the $\Gamma$ point for large vacuum gaps.} of the real-space wave functions from the substrate and the tip are shown, which are directly propagated from localized-basis DFT calculations. The ordering of these wave functions is determined by the magnitude of the eigenvalues of the partial spectral functions of the semi-infinite leads \cite{Paulsson2007}, i.e., the eigenvalues are proportional to the local density of states in the leads. The number of eigenvalues equals the number of bands that cross the Fermi energy. In general, the magnitude of such an eigenvalue is not related to the amplitude of the corresponding wave function in the vacuum region, and thereby not related to the transmission probability for a specific tip-sub combination. This feature is evident in \Figref{fig4}(a) and \Figref{fig4}(b), where the amplitude is seemingly random with respect to the numbering of the wave functions. In addition, unnecessary many wave functions with similar symmetries in the lateral plane are obtained, which is here clearly observed for the single-atom Cu tip, \Figref{fig4}(b), where numerous $s$- and $p$-type wave functions are observed. 

When performing a unitary transformation of these wave functions, the ordering approximately becomes proportional to the maximum amplitude of the wave function, as well as their significance in the tunneling probability for a specific tip-sub combination. This is visualized in \Figref{fig4}(d) and \Figref{fig4}(e), and clearly reflected in the transmission matrix, \Figref{fig4}(f), where large transmission coefficients are observed primarily for the low-number wave functions from each side. The magnitude of a specific element in the transmission matrices is defined as the total transmission of the given tip-sub combination, i.e., after the tip scanning has been performed. Furthermore, wave functions with the same symmetry in the lateral plane are bundled together to give a single wave function with the same symmetry. For instance, the most pronounced unitary transformed Cu-tip wave function, \Figref{fig4}(e), is one single $s$ wave, which should be compared to the normal Cu-tip wave functions, \Figref{fig4}(b), where several wave functions with this symmetry are present. A cutoff is used in the plot range for the wave functions in \Figref{fig4}, in order to highlight also the less important wave functions. For instance, the doubly degenerate lateral $p$ waves from the tip are essentially unimportant in respect to the STM contrast, see \Figref{fig5}(e) and \Figref{fig5}(f). 

As in the case of a single CO molecule adsorbed on a Cu(111) surface, we have previously shown that the non-zero amplitude of certain substrate wave functions away from the molecule is crucial to explain the conductance dip over the molecule \cite{Gustafsson2016,Gustafsson2017}. That is, if a wave function over the substrate has a significant (constant) value that yields a large conductance even when the tip is laterally placed far away from the molecule, this might give a conductance dip over the molecule, as in the case with the CO monomer on the Cu(111) surface. This sign change in the amplitude is evident also for the present CO dimer upon noticing the non-zero amplitude of the first two unitary transformed substrate wave functions [\Figref{fig4}(d), wave function 0 and 1], which, by interference, give rise to a large conductance also when the tip is not centered over the molecule. This feature is the main reason to the slightly elongated conductance dip in the STM image [\Figref{fig2}(b)], whereas its central bright spot is assigned to the large amplitude centered in the first unitary transformed wave function of the substrate [\Figref{fig4}(d), wave function 0].

\subsection{Reduction of tunneling channels}

\begin{figure}[t!]
	\includegraphics[width=\columnwidth]{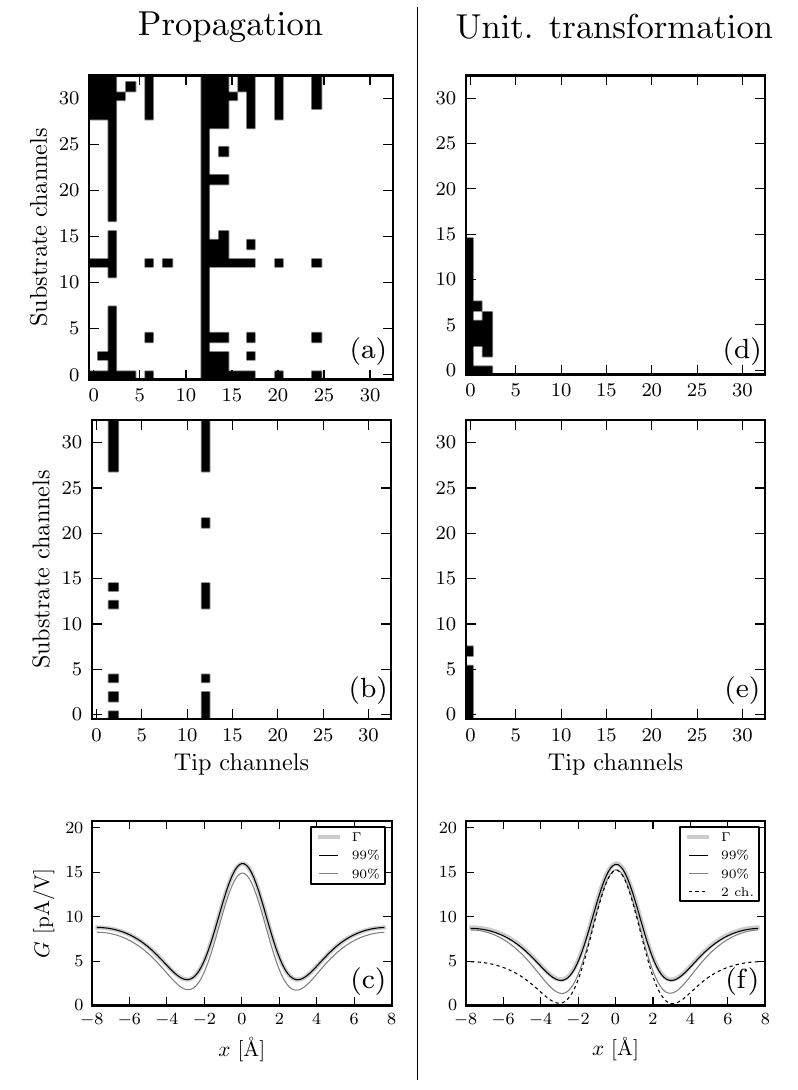}
	\caption{Illustration of the reduction of the number of tip-sub combinations by means of a unitary transformation. Left panel shows (a) [(b)] the wave function combinations (black squares) needed to give an STM image with total conductance that is 99\% [90\%] of the $\Gamma$-point STM image, and (c) shows the cross-section conductance when using these combinations. The right panel shows the corresponding results after the unitary transformation is performed. The dashed line in (f) displays the cross-section conductance when using only two tunneling channels: $\varphi_t^0$, and $\varphi_s^{0\&1}$. \label{fig5}}
\end{figure}

Another important characteristic of a unitary transformation of the wave functions, is that it enables to decrease the number of necessary tip-sub combinations that must be included to recover the STM contrast, as a consequence of the ordering by the amplitude in the vacuum gap for wave functions of the same symmetry. This is demonstrated in \Figref{fig5}. 

We compare to the correct \footnote{That is, by using all tip-sub combinations.} $\Gamma$-point STM image by introducing a cutoff conductance, so that, upon defining X\%-accuracy, the chosen cutoff value gives X\% of the total conductance of the correct $\Gamma$-point STM image. For instance, when using the ordinary wave functions [\Figref{fig4}(a) and \Figref{fig4}(b)], one needs 152 tip-sub combinations [14\%  out of the 1089 ($=33\times33$) combinations], to obtain a 99\% accuracy. Achieving the same accuracy with the unitary transformed wave functions [\Figref{fig4}(d) and \Figref{fig4}(e)] only requires 26 combinations [2.3\% out of 1089 channels]. In the latter case, the majority of the conductance is carried by one single $s$-type tip wave function, as expected by a single-atom Cu tip. 

Reducing the number of channels further (by increment of the cutoff conductance), by compromising slightly with the quality of the STM image [\Figref{fig5}(b) and \Figref{fig5}(e)] shows the same tendency, where the conductance is solely carried via the Cu-tip $s$ wave. Notice that such a calculation  results in an STM contrast that could equivalently be obtained by the Tersoff-Hamann approximation \cite{Tersoff1985}, due to the pure $s$-wave character of the tip. By using an even larger cutoff, so that only two tip-sub combinations are used [tip wave function 0 ($s$ type) \Figref{fig4}(e), and two substrate wave functions, 0 and 1, \Figref{fig4}(d)], may reproduce the $\Gamma$-point STM image qualitatively, so that the origin of the STM contrast is easy to interpret; see dashed line in \Figref{fig5}(f). The latter calculation, however, only gives a 59\% accuracy, which originates from significantly less contribution from the Cu surface. 

\section{Summary}
We have shown that a unitary transformation, by means of the method of Lagrange multipliers, offers a simplified picture when resolving the origin of the STM contrast in terms of individual tip-substrate wave function combinations in the middle of the vacuum region. The unitary transformed wave functions become nicely ordered by (i) their amplitude in the vacuum region, which (ii) approximately determine their importance in the tunneling process, and (iii) are merged into single versions of wave functions of similar symmetry in the lateral plane. The method significantly reduces the number of tunneling channels, and thereby opens up the possibility to make an intuitive and detailed analysis of individual wave function combinations of a specific lateral symmetry. We have further presented an alternative simple formula originating from the Green's function formalism, to describe the transmission probability accounting for multiple-tip-state STM modeling when considering real-space wave functions in the vacuum gap.

\section{Acknowledgements}
The computations were performed on resources provided by the Swedish National Infrastructure for Computing (SNIC) at Lunarc. A.G. and M.P. are supported by a grant from the Swedish Research Council (621-2010-3762).


\end{document}